\documentclass[9pt,twocolumn,twoside]{opticajnl}
\journal{opticajournal} 
\usepackage{soul}
\setboolean{shortarticle}{true}

\usepackage{lineno}

\title{Deployment of a Transportable Yb Optical Lattice Clock}

\author[1,$\dagger$]{Tobias Bothwell}
\author[1,2,$\dagger$]{Wesley Brand}
\author[1,2]{Robert Fasano}
\author[3]{Thomas Akin}
\author[4]{Joseph Whalen}
\author[1,2]{Tanner Grogan}
\author[1,2]{Yun-Jhih Chen}
\author[1,5]{Marco Pomponio}
\author[1,2]{Takuma Nakamura}
\author[6]{Benjamin Rauf}
\author[6]{Ignacio Baldoni}
\author[6,7]{Michele Giunta}
\author[6,7]{Ronald Holzwarth}
\author[1]{Craig Nelson}
\author[1]{Archita Hati}
\author[1,5]{Franklyn Quinlan}
\author[1]{Richard Fox}
\author[3]{Steven Peil}
\author[1,2,5*]{Andrew Ludlow}

\affil[1]{National Institute of Standards and Technology, 325 Broadway, Boulder, Colorado 80305, USA}
\affil[2]{Department of Physics, University of Colorado, Boulder, Colorado 80309, USA}
\affil[3]{Precise Time Department, United States Naval Observatory, 3450 Massachusetts Avenue N.W., Washington, DC 20392, USA}
\affil[4]{Computational Physics, Inc., 8001 Braddock Road, Suite 210, Springfield, VA 22151, USA }
\affil[5]{Electrical, Computer \& Energy Engineering, University of Colorado, Boulder, Colorado, USA}
\affil[6]{Menlo Systems GmbH, Bunsentraße 5, 82152 Planegg, Deutschland}
\affil[7]{Max-Planck-Institut für Quantenoptik, Hans-Kopfermann-Straße 1, 85748 Garching bei München, Deutschland}
\affil[$\dagger$]{Equal Contributions}
\affil[*]{andrew.ludlow@nist.gov}

\begin{abstract}

We report on the first deployment of a ytterbium (Yb) transportable optical lattice clock (TOLC), commercially shipping the clock 3,000 km from Boulder, Colorado to Washington DC. The system, composed of a rigidly mounted optical reference cavity, atomic physics package, and an optical frequency comb, fully realizes an independent frequency standard for comparisons in the optical and microwave domains. The shipped Yb TOLC was fully operational within 2 days of arrival, enabling frequency comparisons with rubidium (Rb) fountains at the United States Naval Observatory (USNO). To the best of our knowledge, this represents the first deployment of a fully independent TOLC, including the frequency comb, coherently uniting the optical stability of the Yb TOLC to the microwave output of the Rb fountain.

\end{abstract}

\setboolean{displaycopyright}{false} 

\begin{document}

\maketitle

The last decade has seen remarkable progress in optical frequency standards (OFS). In laboratory based optical clocks, fractional frequency systematic uncertainties at the $10^{-18}$ level and below have been demonstrated in systems based on both trapped ions \cite{brewer2019al+,zhiqiang2023176lu+,sanner2019optical} and neutral atoms \cite{ushijima2015cryogenic,mcgrew2018atomic,bothwell2019jila,aeppli2024clock}. In optical lattice clocks (OLCs) fractional frequency instabilities below $10^{-16}$ at one second have been realized \cite{schioppo2017ultrastable,oelker2019demonstration}, demonstrating accuracy and precision in OLCs orders-of-magnitude beyond current microwave standards. With frequency ratios between optical atomic clocks reaching uncertainties below $10^{-17}$ \cite{boulder2021frequency}, a redefinition of the second from microwave to optical has become a viable goal \cite{dimarcq2024roadmap}.

\begin{figure*}[h!]
	\centering\includegraphics[trim={0 0.1cm 0 0cm},clip,width=.99\linewidth]{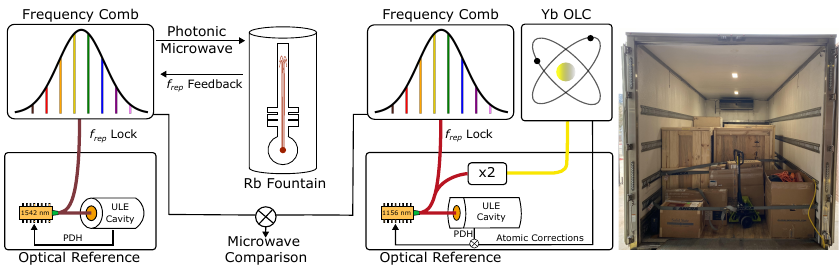}
	\caption{\textit{Left}--Schematic of the USNO Rb/Yb TOLC campaign. The Rb fountain employed enhanced stability be improving LO performance via photonically-generated microwaves. \textit{Right}--Picture of the Yb TOLC and accessories boxed and loaded for transport.}
	\label{fig_1}
\end{figure*}

Despite such marked progress, comparison of optical clocks between geographically isolated locations remains a major challenge \cite{clivati2018optical,pizzocaro2021intercontinental}. While frequency transfer via fiber networks \cite{xu2018studying,cantin2021accurate,schioppo2022comparing} and optical two-way time-frequency transfer \cite{bodine2020optical} have made major advances, such techniques remain limited in applicability and scope. Full verification of optical frequency standards requires transportable clocks, connecting geographically remote metrological institutes and therefore enabling global frequency comparisons. Towards this, PTB demonstrated the first Sr TOLC in 2017 \cite{koller2017transportable}, for the first time showing OLC operation outside the lab. A major breakthrough was reported in 2020 when two transportable Sr OLCs were compared with an elevation difference of 450 m, not only providing a proof-of-principle test of general relativity but showing operation outside the lab into the 18th decade \cite{takamoto2020test}. The year 2020 also marked the first demonstration of a transportable Ca$^+$ clock \cite{huang2020geopotential}. These efforts have shown that clock operation is ready to move beyond the laboratory, with prospects for metrology, geodesy, and fundamental physics.

In this letter, we report on the first deployment of a Yb TOLC. In March of 2023, the system was shipped nearly 3,000 km from NIST in Boulder, Colorado to the United States Naval Observatory in Washington DC for comparisons with the Observatory's Rb fountain clocks (Fig.~\ref{fig_1})~\cite{peil2017microwave}. All components of an independent Yb OLC frequency standard were deployed: atomic physics packages, a frequency comb with microwave output, and a prototype optical cavity with supporting clock laser system. Each component was boxed in a wooden shipping crate, with the full system shipped via a commercial shipping service. This TOLC form factor was chosen to enable future inter-continental deployments towards frequency comparisons with remote national metrological institutes (NMIs).

\textit{Atomic Physics Package}--Central to the Yb TOLC is the integration of the traditional atomic physics system into 3 commercial server racks of approximately 120 cm in height. A forthcoming publication will fully detail the atomic system, including a full systematic uncertainty evaluation \cite{wesley2024}. Briefly, the 3 racks contain a miniature version of the NIST laboratory based Yb OLCs \cite{mcgrew2018atomic}, including a custom vacuum chamber designed for shielding of black-body radiation \cite{fasano2021transportable,wesley2024}. A compact effusive oven combined with a single 100 mW 399 nm diode laser enables loading of $\approx 10^6$ atoms into a broad dipole allowed magneto-optical trap (MOT). A second narrow-line MOT using frequency doubled light from an 1112 nm fiber laser system further cools the atoms to 20 $\mu$K, enabling loading into a 1D magic wavelength optical lattice at 759 nm. The retroreflected lattice uses 1 W from a frequency-filtered commercial tapered amplifier system \cite{fasano2021characterization} and supports Lambe-Dicke confinement of ultracold Yb in trap depths up to 150 lattice recoil energies. A compact 1388 nm laser is used to repump excited clock state atoms to the ground state for readout. All hardware for controlling the physics packages, including  supporting electronics and the control computer, are built into the racks.

\textit{Clock Laser}--To probe the ultranarrow clock transition we  developed a prototype optical cavity which serves as the phase coherent local oscillator (LO) of the Yb TOLC. The cavity, inspired by the cubic design \cite{webster2011force}, realizes an extended  cuboid shape with cut vertices and vent holes specifically engineered to minimize acceleration and holding force sensitivity \cite{fasano2021transportable}. The 10 cm long spacer is built with ultra-low expansion (ULE) glass and uses crystalline mirrors for reduced thermal noise \cite{cole2013tenfold}. The cavity is rigidly held within an Invar cage by four vacuum-compatible plastic balls, enabling robust mounting. Acceleration sensitivities at the mid $10^{-10}$/g (fractional frequency per standard gravity) or better were measured via a 2g flip-over test~\cite{webster2011force}.

\begin{figure*}[t!]
	\centering\includegraphics[trim={0 0.3cm 0 0cm},clip,width=\linewidth]{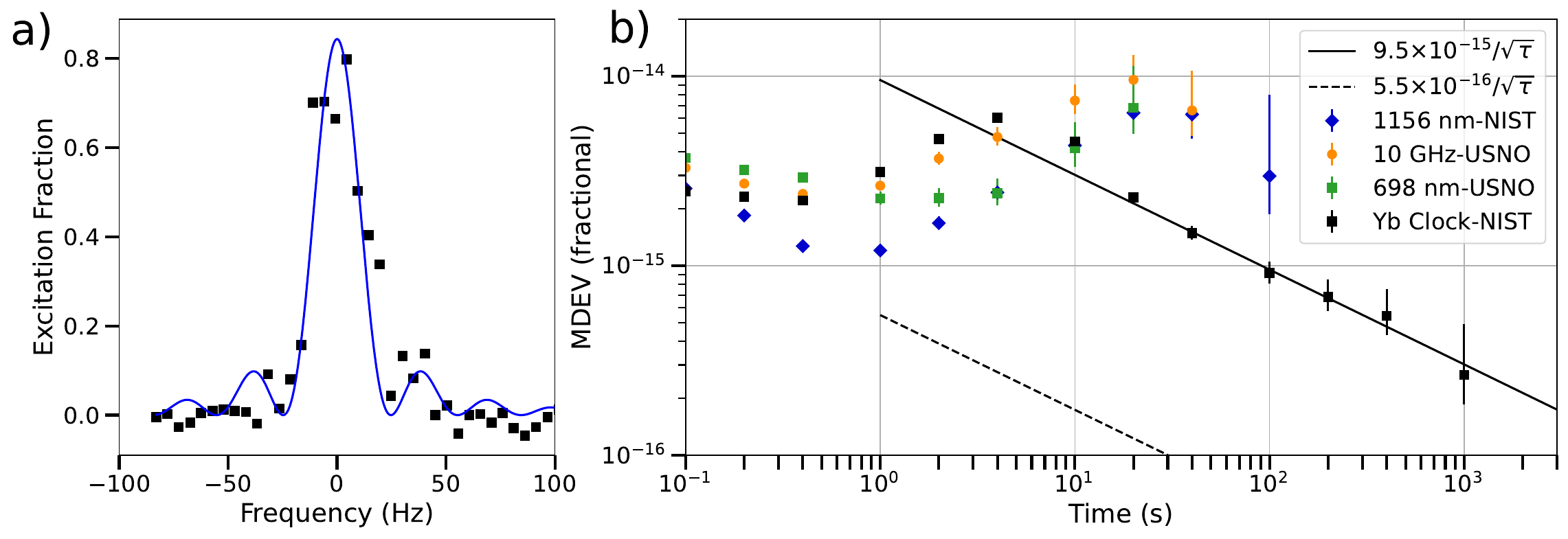}
	\caption{Yb TOLC frequency instability with transportable optical cavity reference. \textbf{a)} Rabi lineshape for a 35 ms $\pi$-pulse with no drift cancellation, taken at USNO after shipping. \textbf{b)} Modified Allan deviations (MDEVs) of the optical cavity stability before and after shipping, as well as full Yb TOLC as measured by laboratory based clocks at NIST. The solid (dashed) black line indicates locked TOLC instability utilizing the transportable (lab-based) LO. Blue, orange, and green points were taken by direct comparison at 1156 nm, 10 GHz, and 698 nm respectively.}
	\label{fig_2}
\end{figure*}

For rapid development of our prototype cavity we used a commercial vacuum chamber without thermal shielding. The ULE cavity was found to have a zero-crossing of thermal expansion near 22 $^\circ$C, necessitating temperature control using Peltier cooling. Despite operation at the zero-crossing, we found the thermal control of the cavity to be the limiting systematic for stability, explained by the prototype's lack of in-vacuum thermal shielding. For deployment the full cavity assembly and supporting lasers for clock light (including frequency doubling of a commercial 1156 nm system) were combined into a single assembly. As shown in Fig.~\ref{fig_2}, the  prototype cavity supported 35 ms Rabi $\pi$-pulses and regularly exhibited instabilities in the low to mid $10^{-15}$ level at timescales relevant for clock operation.

The full Yb TOLC system was tested against the lab-based Yb OLCs at NIST \cite{mcgrew2018atomic}. The optical cavity, with performance as indicated in Fig.~\ref{fig_2}, was steered by the TOLC realizing a fully independent system to compare against the laboratory Yb OLC. Comparison with the lab based Yb OLC showed a frequency instablity of $9.5\times10^{-15}/\sqrt{\tau}$, nearly a factor of 10 lower instability than the USNO Rb fountain. Utilizing a lab-based optical cavity, we observed a significantly improved instability of $5.5\times10^{-16}/\sqrt{\tau}$, reaching long term instability at the low-$10^{-18}$ level. We anticipate improved instability from forthcoming upgrades to the optical cavity and physics packages \cite{wesley2024}.

\begin{figure*}[t!]
	\centering\includegraphics[trim={0 .3cm 0 0cm},clip,width=.99\linewidth]{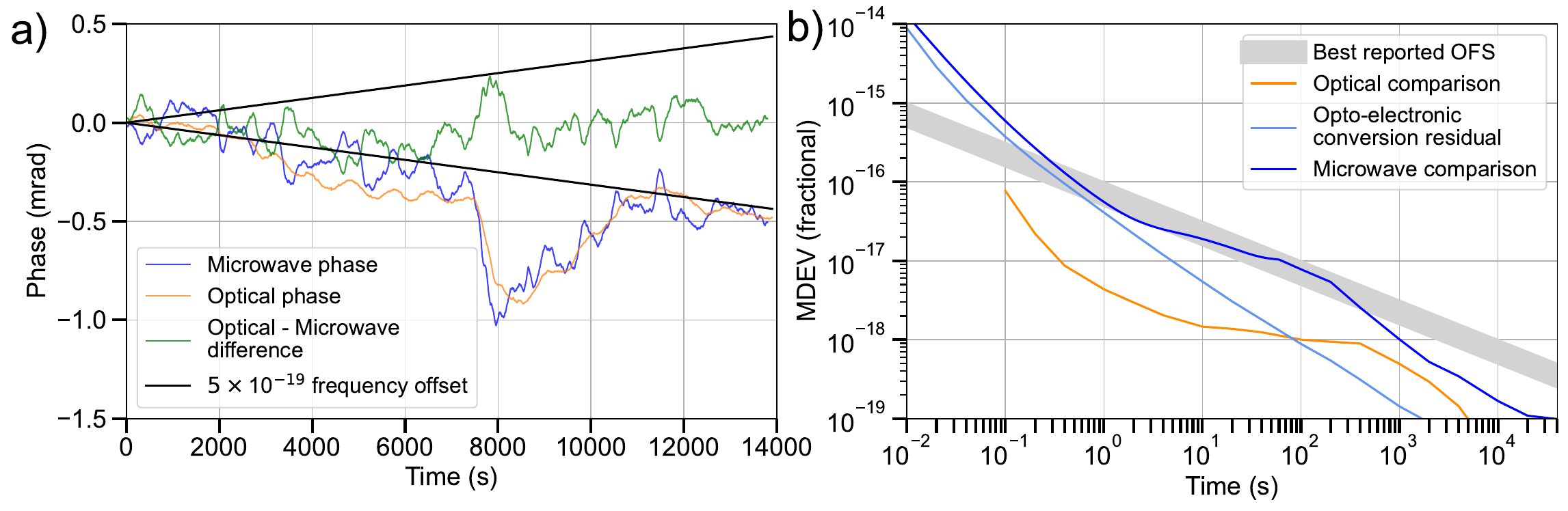}
	\caption{Characterization of optical frequency comb performance \textbf{a)} Phase Evolution of the microwave (blue) and optical (orange, scaled to the microwave frequency) comparisons as well as their difference (green). The cone traced in black represents the boundaries at each given time for the required phase accumulation to cause a fractional frequency offset corresponding to $5\times10^{-19}$ for the 10 GHz signal. \textbf{b)} Modified Allan deviation depicting:  (orange) the optical comb-comb comparison taken at 698 nm; (light blue) the residual phase noise of the Microwave extraction unit at 10 GHz; (blue) the microwave comb-comparison signal stability, representing the absolute upper limit of the optical-to-microwave link stability.}
	\label{fig_menlo}
\end{figure*}

\textit{Comb System}--The Yb TOLC includes a complete comb-based laser reference system \cite{Hansel2017, Giunta2020b}, necessary for realizing a portable OFS. Included within a meter tall rack system are three sub-elements: an optical frequency comb, a spectral purity transfer unit (SPTU)  coherently connecting three distinct clock transition frequencies (1156 nm for doubling to the Yb $^1$S$_0 \rightarrow ^3$P$_0$ transition, 871 nm for doubling to the Yb\textsuperscript{+} $^2$S$_{1/2} \rightarrow ^2$D$_{3/2}$ transition, and 698 nm for the Sr $^1$S$_0 \rightarrow ^3$P$_0$ transition) with an ultrastable 10 GHz photonic microwave signal, and an optical detection unit for the synthesis of comb lines for frequency stabilization of additional Yb specific lasers (1388, 759, and 1112/556 nm). The comb may be stabilized to any of the SPTU frequencies or at 1542 nm.

Before deployment, the phase of both the optical and microwave output signals  (at 698 nm and 10~GHz) between the device utilized in this work and a similar reference comb system were compared on a dead-time free phase-frequency counter while both combs were phase-locked to the same ultrastable cavity-stabilized laser at 1542 nm. The residual phase difference between the optical and microwave traces is shown in Fig.~\ref{fig_menlo}a, where the black cone indicates the phase evolution limits corresponding to an accumulated frequency offset of only $5\times10^{-19}$. The residual of the optical and microwave phase trace has a root mean square value of $101$ $\mu$radians, corresponding to a phase time error of $1.61$ fs. The optical fractional frequency instability (Fig.~\ref{fig_menlo}b) reaches values of $<5\times10^{-18}$ at 1 s and the low-$10^{-19}$ level at a few thousand seconds. This measurement is limited from tens to hundreds of seconds by uncompensated Doppler noise originating in the differential optical path (in fiber) present in the measurement setup. The blue trace in Fig.~\ref{fig_menlo}b reveals the stability of the 10 GHz microwave comparison in contrast to the optical comb comparison at 429 THz (orange). The microwave fractional stability at short term is limited by the residual noise of the opto-electronic conversion in the microwave conversion unit, which provides a limit of $4\times10^{-17}/\tau$ arising from residual flicker phase noise.
The microwave comparison shown here yielded a relative accuracy of $0.43\pm9.87\times10^{-20}$ (Fig.~\ref{fig_menlo}b).

\begin{figure*}[t!]
	\centering\includegraphics[trim={0 0.25cm 0 0cm},clip,width=\linewidth]{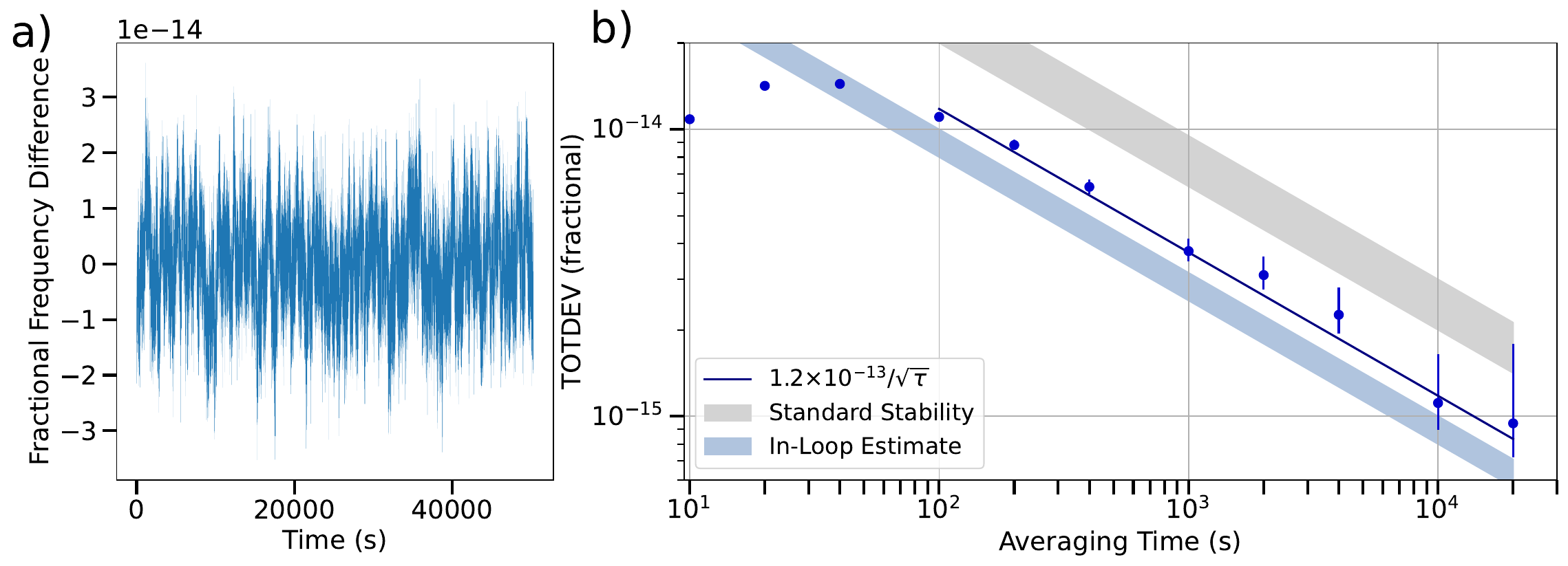}
	\caption{Rb Fountain-Yb TOLC comparison. \textbf{a)} $14$ hour long frequency comparison between the Yb TOLC and USNO Rb fountain at 10 GHz. \textbf{b)} Total Allan deviation (TOTDEV) in fractional frequency units for the comparison data. The Yb TOLC resolved the instability of the Rb fountain to be $1.2\times10^{-13}/\sqrt{\tau}$, reaching the high $10^{-16}$ level after $10^4$ s of averaging.}
	\label{fig_4}
\end{figure*}

\textit{Deployment}--With the full performance of the Yb TOLC characterized the system was ready for deployment. To prepare for shipping the major system components (server rack, cavity, and comb) were each packaged in their own commercial shipping crate. Battery power kept vital systems online (vacuum pumps and temperature controllers) while the system was moved into a commercial shipping truck, at which point power was restored. The procedure was reversed upon arrival at USNO, moving the TOLC into a lab adjacent to the USNO ultrastable laser.
During shipping, a thunderstorm led to a temporary loss of power to the boxed Yb TOLC. The cavity's vacuum was compromised, requiring vacuum pumping upon arrival. Further, the full clock laser assembly suffered several structural failures arising from vibrations loosening screws during shipping. Despite these complications, the cavity and optics remained fully aligned. Within 1 day of arrival all repairs were made and the cavity was found to still be fully aligned to the TEM00 mode, requiring no realignment. The cavity stability against the USNO ultrastable laser is shown in Fig.~\ref{fig_2}b, demonstrating consistent performance before and after shipping. The physics packages arrived free of harm - these racks had been previously shipped and modified to prevent damage during shipping. 

Within 2 days the entire Yb TOLC was fully operational, with clock spectroscopy measuring the optical cavity resonance to have shifted approximately $\approx200$ kHz over 8 days, largely explained by the cavity drift rate. This provided a critical verification of the transportable cavity architecture, demonstrating that standard broadband clock spectroscopy was capable of locating the clock transition after transport. With the full TOLC operational the clock was locked using Rabi $\pi$-pulse times ranging from 8 to 35 ms, observing similar excitation performance and locking abilities as before shipping. 

\textit{Rb Fountain Comparison}--The USNO fountains have been operating nearly continuously for over a decade~\cite{peil2017microwave}, with microwave LO limited instability. To improve short term instability, an improved microwave LO based on down-conversion of an optical LO had been configured to feed one of the Rb fountains, targeting instability $\approx1\times 10^{-13}/\sqrt{\tau}$. To verify the improvements to fountain stability, comparisons between the Yb TOLC comb and USNO comb at both 698 nm and 10 GHz were performed (Figs.~\ref{fig_2}b and \ref{fig_4}). Fountain operation was optimized during the first days of deployment, culminating in an extended overnight comparison as shown in Fig.~\ref{fig_4}a, realizing 14 hours of continuous uptime. The Yb TOLC/Rb comparison showed an instability of $1.2\times10^{-13}/\sqrt{\tau}$, revealing the improved stability of the combined Rb fountain/optical LO system. It further demonstrated the requisite uptime of the Yb TOLC, critical for future frequency comparisons.

In conclusion, we have demonstrated the first TOLC based on Yb, shipping a fully independent Yb OFS nearly 3,000 km. The deployment validated the atomic physics package architecture, with a full accuracy evaluation forthcoming \cite{wesley2024}. We deployed a new optical cavity design \cite{fasano2021transportable}, an important step towards future instability upgrades. Finally, the Yb TOLC provided direct measurement of an optical LO enhanced Rb fountain's stability.

\begin{backmatter}
\bmsection{Funding} 

This work was supported by NIST, DARPA, ONR, and NSF QLCI Award OMA-2016244. T.B. acknowledges support from the NRC RAP.

\bmsection{Acknowledgments} We acknowledge manuscript review by R. Brown and B. Hunt as well as insightful conversations and technical contributions from the NIST Yb team, the Menlo team, A. Roth, and W. Wendler.

\bmsection{Disclosures} The authors declare no conflicts of interest. No US government endorsement of any company is suggested/implied in this work.

\bmsection{Supplemental document}
See Supplement 1 for supporting content. 

\end{backmatter}


\bibliography{references}

\bibliographyfullrefs{sample}

\section*{Fountain Driven by an Optical Oscillator}

The rubidium (Rb) fountains at the United States Naval Observatory (USNO) have operated nearly continuously for over a decade. The clock transition in $^{87}$Rb is driven with a microwave signal derived from an ovenized quartz crystal oscillator. The oscillator’s 5 MHz output is multiplied to 100 MHz, and the 68th harmonic is used to phase lock a 6.8 GHz dielectric resonator oscillator (DRO). The output of a direct digital synthesizer (DDS) is mixed with the 6.8 GHz to reach resonance; the frequency of the DDS is steered to keep the drive centered on the atomic resonance. Typical fountain stability for this configuration is characterized by a white-frequency noise level of $(2-3)\times 10^{-13}/\sqrt{\tau}$ , limited in large part by the phase noise of the quartz oscillator. The quantum-projection noise of the fountains is below $10^{-13}$, reaching as low as $5\times 10^{-14}/\sqrt{\tau}$ for the higher performing systems. By incorporating a photonic oscillator instead of the quartz crystal in the frequency chain, fountain performance can be improved and more closely approach the quantum-projection noise limit. To this end we employ a commercial optical oscillator constructed from telecom wavelength components. A 1542 nm external cavity diode laser is stabilized to a high-finesse cavity made from an ultra-low expansion spacer and silica-substrate mirrors with crystalline coatings to reduce the noise caused by Brownian motion. This system results in a sub-Hz linewidth and an optical stability below $10^{-15}$ at 1 s and an average frequency drift of 3 kHz/day. 

A fiber frequency comb centered at 1550 nm and with a repetition rate of 250 MHz is used to transfer the low phase-noise optical signal to RF. The repetition rate of the comb is
stabilized by phase-locking a comb line to the cavity-stabilized laser at an offset frequency set by an adjustable DDS. An SNR of 40 dB in a 100 kHz bandwidth enables the phase stability of the laser to be transferred to the frequency comb to a high degree. Two different optoelectronic paths are used to generate microwave and RF signals. In the first path, the 4th harmonic of the frequency comb at 1 GHz is generated using a conventional photodetector, bandpass filtered and divided to produce 200 MHz. This 200 MHz signal is used as the reference for the DRO in a fountain’s microwave chain, as well as being divided to 5 MHz and input to our clock measurement system, which measures the ensemble of atomic clocks that contribute to the USNO timescale. A second path uses a three-stage interleaver to multiply the repetition rate to 2 GHz, and the 5th harmonic at 10 GHz is extracted from a high speed photodetector. This 10 GHz signal preserves the optical stability to a high degree. The optical oscillator’s frequency drift intersects the fountain’s integration at about 40 s, at which point the drift of the optical cavity is disciplined by the rubidium fountain. To steer the repetition rate of the frequency comb, we feed back to the DDS fixing the offset frequency of the heterodyne beat-note between the cavity-stabilized laser and the optical frequency comb. The hybrid clock exhibits optical-level frequency stability at short averaging times and integrates as a fountain with white-frequency noise level of $1.2\times 10^{-13}/\sqrt{\tau}$, reduced compared to using a quartz oscillator, as shown in Fig.~\ref{fig_4}b.

\end{document}




\preprint{}
\title{Supplementary Material for Lattice Light Shift Evaluations In a Dual-Ensemble Yb Optical Lattice Clock}
\author{Tobias Bothwell}
\affiliation{NIST}
\author{Ben}
\affiliation{NIST}
\author{Jacob}
\affiliation{NIST}
\author{Party}
\affiliation{NIST}
\author{Andrew Ludlow}
\affiliation{NIST}
\maketitle

\section{Fitting Sideband Spectra to Extract $<n_z>$}
As the method to prepare one ensemble in $n_z=0$ and the other in $n_z=1,2$ does not prepare $n_z=1,2$ with perfect fidelity, we use sideband spectroscopy to extract the $<n_z>$ of the ensemble. Fitting the sidebands to a standard harmonic treatment \cite{blatt2009sideband} does not yield accurate results because our $n_z$ populations are non-thermal. Additionally, with Sisyphus cooling the individual axial states become resolved, revealing the harmonic model does not capture the corner frequency of higher order axial states. 

To address these concerns we utilize a model which reproduces the experimental data. We still find the optical trap depth as found from the corner frequency of the harmonic model. To correctly capture higher order band spacings we then evaluate each bands corner frequency using Mathieu functions as outlined in \cite{beloy2020modeling}, reproducing the observed spacing. We then simulate the radial temperature's smearing of the sideband lineshape by sampling an extended series of frequencies below each $n_z$ corner frequency, approximating a radially averaged effective trap depth \cite{nemitz2019modeling}. A radial temperature is then used to provide a Boltzman weighting to each effective trap depth within a single $n_z$ state. Finally, a Lamb-Dicke parameter is also associated with each sampled point using the effective trap depths' frequency as input. Note that this radial modeling is performed for each $n_z$ state separately, but with a single radial temperature.

To model each lineshape, we assume a time averaged Rabi lineshape. For a specific clock frequency, each sampled point has a unique detuning and Lamb-Dicke parameter, adjusting the effective Rabi frequency \cite{leibfried2003quantum,ludlow2008strontium}. The simulation thus performs a least-squares fitting routine to optimize agreement between our model and data. Fit parameters are the weights for $n_z$ states, Rabi frequency, and radial temperature. We find this model reproduces the observed sideband spectra to an excellent degree as shown in Figure \ref{figure_sideband_scan_with_fit}. We assign error from both fitting and disagreement in $<n_z>$ from hand fitting of the amplitudes of the data at the corner frequencies, typically corresponding to an $<n_z>$ uncertainty of $\approx .1$.

\begin{figure}
	\includegraphics[width=0.49\textwidth]{sideband_fit_v1.png}
	\caption{A representative sideband fit (blue line) for ensemble 2 prepared ideally in $n_z=2$ at 123 $E_R$ is shown. The error bars on the data points are the standard deviation of two points for each frequency added in quadrature with a machine specific readout noise of 0.5\%. We see an $n_z=0$ population of 19.2\%, $n_z=1$ of 7.6\% and $n_z=2$ of 73.1\% Not shown is the sideband scan for ensemble 1, ideally prepared in $n_z = 0$, which has a residual population in $n_z=1$ of $8(2)\%$. }
	\label{figure_sideband_scan_with_fit}
\end{figure}

\section{Multi-ensemble Frequency Comparison}

Extracting the frequency difference between multiple ensembles in a single clock from the measured excitation fractions for each ensemble requires converting excitation fraction to frequency. To do so we utilize a Rabi line-shape  \cite{steck2007quantum}

\begin{equation}
	\label{eqn:Rabi}
	P_i=C_i \frac{\Omega^2}{\Omega^2+\delta_i^2}\mathrm{sin}^2\left(\frac{T_\pi}{2}\sqrt{\Omega^2+\delta_i^2}\right),
\end{equation}

\noindent
where $P_i$ is the excitation fraction of the of ensemble $i$, $C_i$ the contrast, $\Omega$ is the Rabi frequency, $T_\pi$ the pulse time optimized for a $\pi$-pulse, and $\delta_i$ the detuning from resonance. For clock operation the Rabi lineshape is probed on both sides of the lineshape with the frequency difference given by the full-width-at-half-max (FWHM). Each ensemble's contrast is then taken as two times the average excitation from an experimental run, capturing in-situ differences in excitation fraction. 

As in standard OLC operation \cite{ludlow2008strontium}, we probe the clock transition via two opposite hyperfine spin states ($m_F = \pm 1/2$) by interrogating both sides of each spin state's Rabi lineshape, dithering the frequency by the Rabi lineshape FWHM. For a single ensemble and single hyperfine state we can then take the difference in excitation fraction as measured on opposite sides of the lineshape and, using Eq. \ref{eqn:Rabi}, map the difference to a frequency offset from the true lineshape center. The difference between this frequency offset between ensembles then provides a frequency difference between ensembles with common-mode laser noise, leading to enhanced stability.  

To bound potential errors from conversion of excitation fractions to frequency, we perform simulations of our analysis using data generated from the analytical line-shape of Eq.~\ref{eqn:Rabi}, including laser noise and quantum projection noise \cite{ludlow2015optical}. We find that this method generates errors in the frequency difference between the ensembles linear in the error in the contrast. For example, a $2\times10^{-17}$ shift measured between $n_z=0$ and $n_z=2$ would incur a $1\times10^{-18}$ error for a 5\% error in the contrast. Measurement of the contrast to $<1\%$ error is evaluated from the average excitation fraction over the entire clock measurement. Even with perfect reproduction of contrast, the simulations suggest potential biasing of results (similar to a servo-error in standard clock evaluations), arising at the level of $<1\%$ of the measured shift. At the 1\% level this bias is only relevant for the large detunings in the running wave experiment of the main text. While we anticipate this bias is largely canceled in the measurement of the running wave magic frequency, we include an additional 1\% uncertainty relative to the measured differential frequency shift for all running wave measurements.

\section{Preparation of Motional State Ensembles}

We use the vertical magnetic field gradient of the MOT coils to spectroscopically resolve ensembles 1 and 2, allowing preparation of ensemble 1 in $n_z\approx0$ and ensemble 2 in $n_z\approx1$ or $2$. We begin by preparing both ensembles in $n_z \approx 0$. The preparation process, outlined below, is shown in Figure \ref{fig:select} for steps \textbf{a)} through \textbf{f)}.
For \textbf{a)}, we apply a magnetic field gradient such that B\textsubscript{vertical} is approximately $B_z=0$ G for ensemble 1, and $B_z \approx 5$ G for ensemble 2. 
\textbf{b)} uses a 1.2 ms $\pi$-pulse to prepare $\approx60\%$ of ensemble 1 in the excited clock state, which is nominally unperturbed by the Zeeman shift, while selecting only a few percent of ensemble 2. 
The 1.2 ms pulse is experimentally chosen to optimize the pulse fidelity for ensemble 1 while minimizing excitation of ensemble 2, largely protected by the large applied Zeeman shift.
Accidental excitation of ensemble 2 results in imperfections in $n_z$ preparation.

With the vertical magnetic field gradient still on, in \textbf{c)} we apply a 20-ms resonant 556-nm pulse to clear the $\approx40\%$ of atoms remaining in the ground clock state of ensemble 1, with ensemble 2 once more shielded by the differential applied magnetic field. 
Ground state atoms in ensemble 1, with approximately no magnetic field, are resonant with the 556-nm transition and are heated out.  We typically see this blow-away pulse is $>97\%$ effective for ensemble 1, while retaining $\approx95\%$ of ensemble 2. The magnetic field gradient is then turned off at the beginning of \textbf{d)} and given 20 ms to settle before an adiabatic rapid passage (ARP) pulse on the first (or second), order blue sideband is applied globally to both ensembles. 
The excited atomic sample in ensemble 1 is in $n_z=0$, a dark state for the frequency of the applied ARP pulse. Ensemble 2 is promoted to the exited clock state and $n_z=1$ ($n_z=2$) depending on excitation on the first (second) blue sideband, with 80\% (25\%) transfer efficiency. 

In \textbf{e)}, any remaining ground clock state atoms in either ensemble are heated out via a 5 ms 399-nm pulse.
Finally for \textbf{f),} ensemble 1 and ensemble 2, both in the excited clock state in $n_z=0$ and $n_z=1$ or $2$ respectively, are moved to the ground state via an ARP pulse on the carrier transition followed by 1388-nm light to repump any atoms not moved by the carrier ARP. 
From here optical pumping is performed before proceeding to differential spectroscopy of the ensembles.

With our current control capabilities, a non-negligible fraction of the final sample of ensemble 2 is found to be in motional states other than desired. 
This is dominated by the accidental selection of a few percent of $n_z=0$ atoms to \textsuperscript{3}P\textsubscript{0} in  \textbf{b)}.
We note that separating the samples by $\sim1$-mm, several times farther distance than in this paper, reduced this accidental selection to $<$1\% suggesting we are limited by off-resonant excitation. A less significant source of motional state impurity is due to the optical pumping after \textbf{f)}, where scattered photons may change the motional state. The selection and spin polarization process do cause some heating of ensemble 1 to $n_z\sim0.13$. We also search for $n_z=3(2)$ population in the $n_z=2(1)$ sample in ensemble 2, via sideband spectroscopy. No excitation was found for the $\Delta n_z=-3(2)$ sideband, so we do not assign any population to $n_z>2(1)$.

\begin{figure*}
	\centering
	\includegraphics[width=0.98\textwidth]{appendixFig_TobyLightShiftPaper_v1.png}
	\caption{The process used to prepare ensemble 1 in $n_z\approx0$ and ensemble 2 in $n_z\approx 1$ or $2$ is shown. Each step outlined in the text is shown sequentially as \textbf{a)} through \textbf{f)} with the length of the step shown at the bottom. Percents denote the fraction of the original sample while yellow arrows are used for 578-nm laser pulses, green for 556-nm laser pulses, and blue for 399-nm scattered photons. $E_{Zeeman}$ denotes the Zeeman shift of the clock transition due to the vertical magnetic field gradient on ensemble 2, shown here for simplicity affecting \textsuperscript{3}P\textsubscript{0} only, and $E'_{Zeeman}$ is the Zeeman shift of the 556-nm transition, again shown only for \textsuperscript{3}P\textsubscript{1} for simplicity. The levels unperturbed by the Zeeman shift are shown with lighter shading. For steps \textbf{d) }through \textbf{f)} either the $n_z=1$ or the $n_z=2$ option is performed depending on the desired final motional state. These final motional state options only differ by performing the ARP pulse in \textbf{d)} on the $\Delta n_z=1$ or $\Delta n_z=2$ sideband. }
	\label{fig:select}
\end{figure*}

\section{Cold Collisions}

The cold collision shift, arising from atomic interactions, requires characterization and correction for the presented results in the main text. A benefit of the evaluation of $\frac{\delta \tilde{\alpha}^{E1}}{\delta \nu}$ via interleaved lattice frequencies is identical preparation and therefore cold collision shifts. The data suggests differences of $<10$ atoms between the two conditions, suggesting cold collision corrections orders of magnitude below the uncertainties presented in Figure 2. Identical atom number preparation and density is not feasible for the differential measurements between ensembles. In the following we discuss cold collision corrections in such measurements.

\subsection{Differential Cold Collision Shifts}

Ratchet loading provides an intuitive method to vary the relative density between two samples of lattice loaded atoms. By varying the efficiency of the shelving pulse, it is possible to obtain large differences in atom number between the two ensembles. After longitudinal state preparation, synchronous spectroscopy of these two samples provides a direct measurement of the differential frequency shift as a function of atom number difference between these samples. While this method provides a rapid method for evaluation compared to interleaved comparisons, it can only provide a correction at the 10\% level due to a small difference in sample length between ensemble 1 and 2.

For Sisyphus cooled samples as used in the differential motional state study (Figure 4) of the main text density shifts were evaluated by measuring the frequency difference between otherwise identically prepared ensembles. Critically, when both ensembles are prepared in the same $n_z$ state the fidelity of preparation is far higher than discussed in the differential $n_z$ preparation. For $n_z =$0, 1, and 2 the cold collisional shift in fractional frequency units of $10^{-18}/1000$ atoms was measured at -6.14 (0.42), -5.52(0.63) and -4.11(0.42) at trap depths of 124, 129, and 135 $E_{rec}$. To apply corrections to the data of Figure 4, each frequency difference had a $<n_z>$ weighted cold collision shift correction based on the fit sideband spectra. Shifts were scaled to the correct depth using a $U^{5/4}$ scaling \cite{nicholson2015new,nicholson2015systematic} due to the challenges associated with measuring cold collisional shifts at shallow trap depths. For thermal samples without sideband cooling a $U^{3/2}$ scaling may hold - additional uncertainty was included to  conservatively account for the difference between these models. The data in Figure 4 is thus well described at large trap depths where cold collision shifts are dominant and where extrapolation of the coefficients is minimal; at shallow depths the applied corrections are vanishingly small. For context, the largest applied cold collision correction in this data is 3.5$\times 10^{-18}$. 

For the auxiliary running wave measurements of Figure 3 the atoms were not Sisyphus cooled as discussed in the text. In this regime the cold collisional shift is suppressed \cite{mcgrew2018atomic}. Additionally care was taken at the deepest depths to ensure $<$200 atom number difference between regions with $\approx 1000$ atoms. Under such conditions we expect that magnitude of  cold collisonal correction to be of order $1\times10^{-19}$, providing an uncertainty on the running wave magic frequency an order of magnitude smaller than quoted in the text. Similarly we estimate a difference in trapping volume between regions as caused by the running wave to be at the 1\% level, negligible for this work.

\section{Running Wave Light Shift}

\bibliography{references.bib}